# The expanded Maxwell's equations for a mechano-driven media system that moves with acceleration

Zhong Lin Wang*

1. Beijing Institute of Nanoenergy and Nanosystems, Chinese Academy of Sciences, Beijing 101400, P. R. China
2. School of Nanoscience and Technology, University of Chinese Academy of Sciences, Beijing 100049, P. R. China
3. School of Materials Science and Engineering, Georgia Institute of Technology, Atlanta, Georgia 30332-0245, USA.

* Corresponding author: zhong.wang@mse.gatech.edu

**Abstract**

In classical electrodynamics, by motion for either the observer or the media, it always naturally assumed that the relative moving velocity is a constant along a straight line (e.g., in inertia reference frame), so that the electromagnetic behavior of charged particles in vacuum space can be easily described using special relativity. However, for engineering applications, the media have shapes and sizes and may move with acceleration, and recent experimental progresses in triboelectric nanogenerators have revealed evidences for expanding the Maxwell's equations to include media motion that could be time and even space dependent. Therefore, we have developed the expanded *Maxwell's equations for a mechano-driven media system* (MEs-f-MDMS) by neglecting relativistic effect. This article first presents the updated progresses made in the field. Secondly, we extensively investigated the Faraday's law of electromagnetic induction for a media system that moves with an acceleration. We concluded that, the newly developed MEs-f-MDMS are required for describing the electrodynamics inside a media that has a finite size and volume and move with and even without acceleration. The classical Maxwell's equations are to describe the electrodynamics in vacuum space when the media in the nearby are moving.

Keywords: Maxwell's equations for mechano-driven slow-moving media, Faraday's law of electromagnetic induction, non-inertia reference frame

## 1. Introduction

Studying the electrodynamics of a moving medium has had a long history that was first started by Maxwell [1]. Hertz systematically extended Maxwell's theory for moving media [2] but his equations were valid only for conductors. Minkowski derived electrodynamic equations for moving media using the principle of relativity [3]. The interest on the study of electrodynamics of moving media has been revived in recent decades [4, 5]. There are a number of studies about the Maxwell's equations (MEs) for moving media/bodies with a focus on the scattering, reflection and transmission of electromagnetic waves from moving media [6, 7, 8]. However, all of the existing studies are usually for describing the electromagnetic phenomenon in the Lab frame as observed by an observer who is moving at a constant velocity along a straight line. The most classical and well received theory is special relativity by using Lorentz transformation. As for generations of students, this is the taken as is classical method for dealing with the electromagnetism of moving medium [9, 10]. In general, physics phenomena should be the same when viewed by two observers moving at a constant relative velocity, e.g., the inertial frames are equivalent with each other. Special relativity is unique for describing the electromagnetic behavior of moving charged particles in the microscopic world and in the universe, and the MEs for vacuum space has served as the foundation for the field theory. However, such beautiful theory is based on an important assumption: there is no acceleration in the reference frame so that there is no input work from external mechanical force, therefore, the total energy is conserved between electricity and magnetism.

In practice, however, most of the phenomena are observed with the media moving even with acceleration, such as circular motion and oscillation etc. Once the movement is in a non-inertia frame, there must be a contribution from an external force that drives the media to move. One may say that we can treat such a case using the general relativity, but the solution of which is so complex and difficult for engineering purposes. On the other hand, there is no absolute inertia reference frame in the universe precisely speaking, because everything is in motion. To develop an effective approach that can describe the electrodynamics of a media system that moves at a varying velocity, we have recently developed the *Maxwell's equations for a mechano-driven media system* (MEs-f-MDMS) [11, 12], in which the media have time-dependent shapes, volumes and boundaries, and more importantly, the movement of the media is described by a velocity field that is time and space dependent, $\boldsymbol{v}(\boldsymbol{r},t)$. The only required condition is that the moving speed is much less than the speed of light ($v$ << c) by ignoring the relativistic effect. The equations are to describe the coupling and interactions among (mechano) force - electric – magnetic fields.

In this paper, we will review the experimental observations that have been made to illustrate the needs for expanding the MEs. Then, we will present the progresses that have been made in developing the MEs-f-MDMS. Finally, we will present the expansion of the Faraday's law of electromagnetic induction for media that are moving with acceleration and lay out the conditions under which the MEs have to be expanded.

## 2. Experimental evidences for developing the Maxwell's equations for a mechano-driven system

The motivation for developing the MEs-f-MDMS is driven by the recent experimental progress made using triboelectric nanogenerators (TENGs) ) [13, 14]. TENG is a new technology that uses the triboelectrification effect, electrostatic induction effect and relative movement of the media for converting mechanical energy into electric power/signal. A physical contact between two dielectric media produces triboelectric charges of opposite signs on their surfaces. A change in spatial distribution of the media as driven by an external force, surface electrostatic charge density, as well as the distance between the two electrodes built on the surfaces of the two media, results in a variation of electric field in space, which is a form of displacement current that generates an output conduction current across the load connected between the two electrodes. In general, the operation frequency of TENG is slow and the moving velocity of the media is rather low, so that one may ignore the electromagnetic radiation due to medium movement.

However, there are some recent experimental progresses that force us to reexamine the electromagnetic behavior of TENGs. First, using a rotation based TENG in sliding mode, it was possible to wirelessly deliver electric power to a reasonable distance using the displacement current produced by media movement [15]. Multiple LED lights were simultaneously lit up using the displacement current. The operation frequency of a MEMS based TENG can even reach over 1.1 MHz, which is high enough for producing electromagnetic radiation [16]. Using the electromagnetic radiation produced by a TENG by a fingertip movement, electric signals can be transmitted to a distance of many meters [17], suggesting the potential of TENG for wireless signal transmission. Recently, using the displacement current produced by triggering a TENG using human voice, it has been demonstrated to transmit wirelessly for a distance over 5-10 m under sea water, showing the possibility of short range wireless communication in water using TENG generated electromagnetic wave [18]. By relatively sliding dielectric media, significant power can be generated wirelessly for lighting and powering electronics [19]. All of these experimental observations show that mechanical triggering of TENG can produce a high enough wireless signal. As a general case, the media boundaries here do vary with time, and therefore, we need to systematically expand the MEs for moving charged media system with acceleration, which could be important not only for quantifying the output of TENG and optimization of device design, but also for predicting future new potential applications and technologies. This is the motivation we started to look into the fundamentals of Maxwell's equations.

## 3. Medium is not an aggregation of charges

The field due to a moving charge in vacuum space can be calculated precisely using the Liénard-Wiechert potential even it has an acceleration, which is correct because the charge is a point without volume and boundary. A stationary charge can be represented by a point, and the instantaneous current produced by a moving charge is represented by a delta function multiplied by its moving velocity. This model has been adopted for calculating the radiation from a moving charge in accelerators.

A medium is not just an aggregation of charges, but composed of atoms and electrons. A solid material is a typical example of media that has certain permittivity, and it usually has a specific shape and size. A medium is an assembly of atoms in specific order and chemistry so that it has certain dielectric, electric and elastic properties, so that it can have different electrical, optical, thermal and mechanical properties. The electromagnetic field due to a moving medium cannot be calculated using the Liénard-Wiechert potential. The full solution of the MEs has to meet the boundary conditions on the medium surfaces.

## 4. Electrodynamics of a moving media system in inertia reference frame

In physics, a reference frame is always chosen for describing the physics laws in mathematical forms, and the inertia frame is the simplest one. The mathematical expression of a physics law should take the same format in all inertia reference frames. In electrodynamics, one reference frame is in the Lab frame where the observer is at rest. The other reference frame is attached to the medium that is moving, which an inertia frame if the medium movement has no acceleration and it is a non-inertial frame if the medium moving moves with an acceleration.

Someone may say that the field due to a moving charge can be calculated precisely using the Li énard-Wiechert potential even it has an acceleration, which is correct because the charge is a point charge without volume and boundary. The moving charge can be viewed as a pulsed current density in space that is represented using a delta function. For a moving medium that has a surface and volume with certain permittivity, the field due to a moving medium cannot be calculated using the Liénard-Wiechert potential. The solution has to meet the boundary conditions on the medium surface. This different must be kept in mind in order to avoid any further misunderstanding. All of our discussions are about medium that has a volume and boundary. A medium is an assembly of atoms in specific order and chemistry so that it has certain dielectric, electric and elastic properties, so that it can have different electrical, optical, thermal and mechanical properties. A medium is NOT just an aggregation of charges!

## 5. The Maxwell's equations and special relativity for a moving observer in an inertia reference frame

To start our discussion, we first start from the integral forms of the four physics laws that are directly mathematical expressions of the experimentally observed physics phenomena [20, 6]:

$$\oiint_S \boldsymbol{D}' \cdot d\boldsymbol{s} = \iiint_V \rho_f \, d\boldsymbol{r}.$$

Gauss's law for electricity (1a)

$$\oiint_S \boldsymbol{B} \cdot d\boldsymbol{s} = 0.$$

Gauss's law for magnetism (1b)

$$\oint_C \boldsymbol{E} \cdot d\boldsymbol{L} = -\frac{d}{dt}\iint_C \boldsymbol{B} \cdot d\boldsymbol{s}.$$

Faraday's electromagnetic induction law (Lenz law) (1c)

$$\oint_C \boldsymbol{H} \cdot d\boldsymbol{L} = \iint_C \boldsymbol{J}_f \cdot d\boldsymbol{s} + \frac{d}{dt}\iint_C \boldsymbol{D}' \cdot d\boldsymbol{s}.$$

Ampere-Maxwell law (1d)

where $\rho_f$ is the density of free charges in space, and $\boldsymbol{J_f}$ is the current density. The surface integrals for $\boldsymbol{B}$ and $\boldsymbol{D'}$ are for a surface that is defined by a closed loop c, and they are the magnetic flux and displacement field flux, respectively. *Note*: we use $\boldsymbol{D'}$ instead of $\boldsymbol{D}$ is to keep consistent with our previous publications, rather than the displacement vector in a moving frame of reference. Eq. (1a) means that the total electric flux through a closed surface is the total charges enclosed inside the surface. Eq. (1b) means that the total magnetic flux through a closed surface is zero. The Faraday's law of electromagnetic induction (Eq. (1c)) is that the changing rate of the *total* magnetic flux through an open surface is the induced electromotive potential around its closed-edge loop. The Ampere-Maxwell law (Eq. (1d)) is that the changing rate of the *total* electric flux through an open surface plus the total current flowing across the surface is the integral of the magnetic field around its closed edge loop. The law of the conservation of charges is:

$$\oiint_S \boldsymbol{J_f} \cdot d\boldsymbol{s} + \frac{d}{dt}\iiint_V \rho_f \, d\boldsymbol{r} = 0. \tag{2}$$

The integral form of the MEs is more general, but the most commonly used MEs are in differential form. Now we make an important assumption: *the volume and shape/boundaries of the dielectric media in space are fixed and at stationary in the Lab reference frame*. Under such an assumption for fixed boundaries, the time differentiation can be directly switched with the integral in Eqs. (1c, 1d), and directly applied onto the corresponding function without considering the change in integrating surface or line. By using the Stokes theorem, we have

$$\nabla \cdot \boldsymbol{D'} = \rho_f \tag{3a}$$

$$\nabla \cdot \boldsymbol{B} = 0 \tag{3b}$$

$$\nabla \times \boldsymbol{E} = -\frac{\partial}{\partial t}\boldsymbol{B} \tag{3c}$$

$$\nabla \times \boldsymbol{H} = \boldsymbol{J_f} + \frac{\partial}{\partial t}\boldsymbol{D'} \tag{3d}$$

$$\nabla \cdot \boldsymbol{J_f} + \frac{\partial}{\partial t}\rho_f = 0 \tag{3e}$$

This is the most familiar form of the MEs. However, one must point out that *the differential form of MEs applies only to the cases in which the volumes and boundaries of the dielectric media are time-independent*.

This is the case if one is only interested in the generation and transmission of electromagnetic waves for stationary media! One has to keep this in mind.

As for moving media, most of the readers must first use special relativity. Special relativity is about the electromagnetic phenomena observed in vacuum space without considering the shape and boundary of the medium, which is the case for moving charged particles in space. Some readers may simply assume that the charges distributed in space are made of point charges without considering their collective property as a matter such as dielectric properties. This is an excellent approximation for describe the electromagnetic behavior in the universe, in which all stars including earth can be considered as points due to the vast space of the universe.

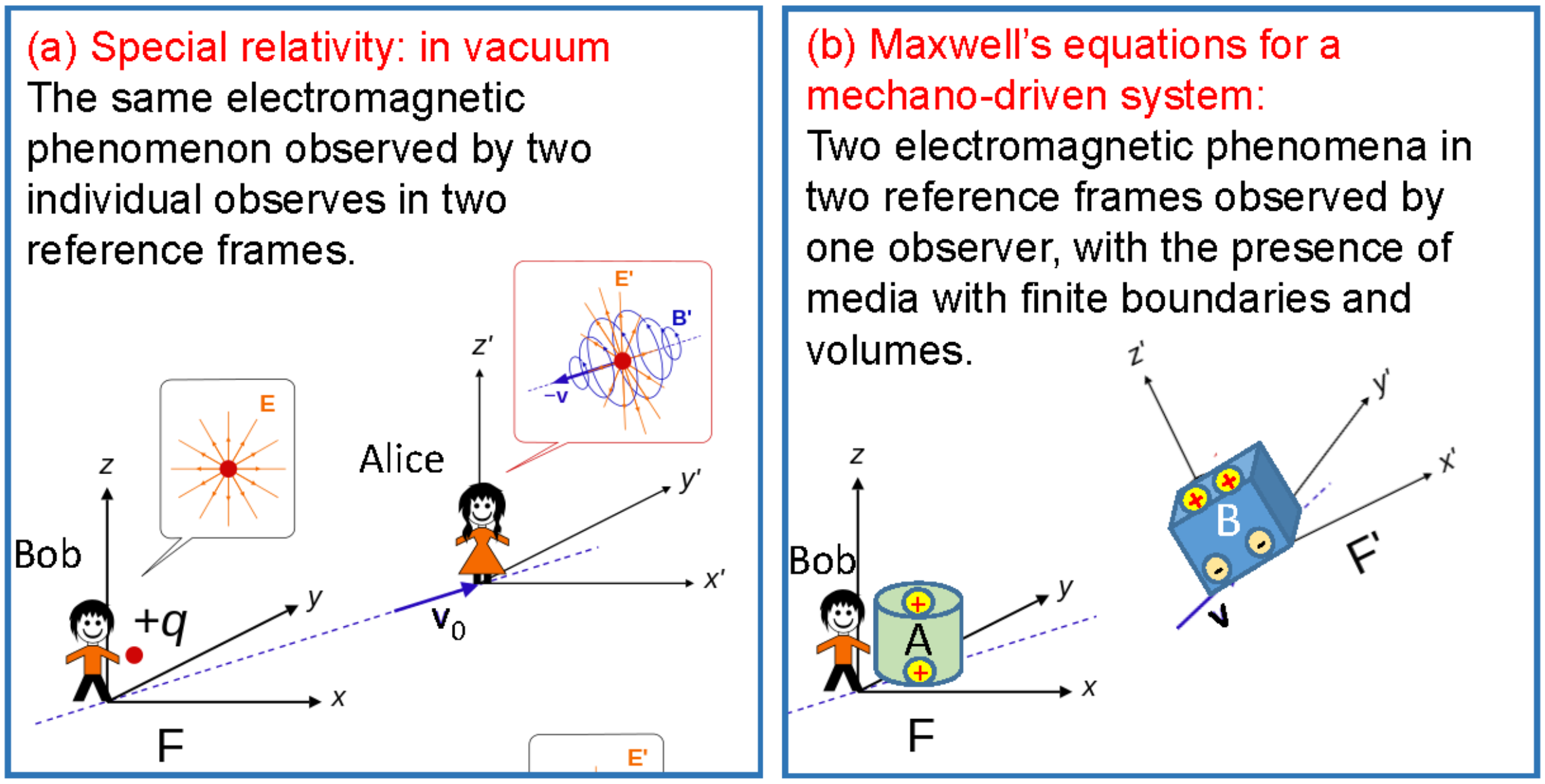

Figure 1. (a) Special relativity is about the experience of two observers, Bob and Alice, who are relatively moving at a constant velocity and along a straight line, about the same electromagnetic phenomenon in vacuum space. In this case, the speed of light in vacuum space is assumed invariant regardless which reference frame it is measured. (b) The MEs-f-MDMS is about the observation of one observer regarding to two electromagnetic phenomena that are associated with two relatively moving media, respectively; the media have size and shape and they may move with acceleration as driven by an external force. The two phenomena may have interaction.

In special relativity, the moving velocity of the observer is a constant $\boldsymbol{v}_0$ and along a straight line, as shown in Fig. 1a, so that the moving reference frame where an observer Alice is at is an inertia reference frame with respect to the rest reference frame where Bob is in (Lab frame). Physical phenomena are the same when viewed by two observers moving relatively at a constant velocity $\boldsymbol{v}_0$, provided the coordinate in the space and time for the two frames are related by the Galilean transformation: $\boldsymbol{r}' = \boldsymbol{r} - \boldsymbol{v}_0 t$, and $t' = t$. The special relativity is about the observation of the same electromagnetic phenomenon by two observers in the two inertia reference frames, respectively, e.g., one phenomenon but two observers. The electromagnetic fields observed by Bob are governed by the MEs given by Eq. (3). The electric and magnetic fields as observed by Bob $(\boldsymbol{E}, \boldsymbol{B})$ and Alice $(\boldsymbol{E}', \boldsymbol{B}')$ can be correlated using the Lorentz transformation. This is an excellent approach for considering the electromagnetic behavior of charged particles in vacuum as observed in different reference frames, but it may have difficult to deal with the electromagnetic behavior of dielectric media that is distributed in a portion of space. In theoretical physics, this is considered as the classical approach used in the field theory, which holds only for inertia reference frames, as labeled as Approach 1 in Fig. 2. The MEs for stationary media were firs derived from the four physics laws. Then the fields observed in a moving inertia frame of reference is given using the Lorentz transformation. There is no restriction on the moving velocity of the reference frame except it is less than the speed of light in vacuum space. Such an approach is most effective for electrodynamics in vacuum space, but it may not be easily applied if there are media, because it is not sure what will be used as the speed of light in the Lorentz transformation if part of the space is occupied by shape and volume specified dielectric media.

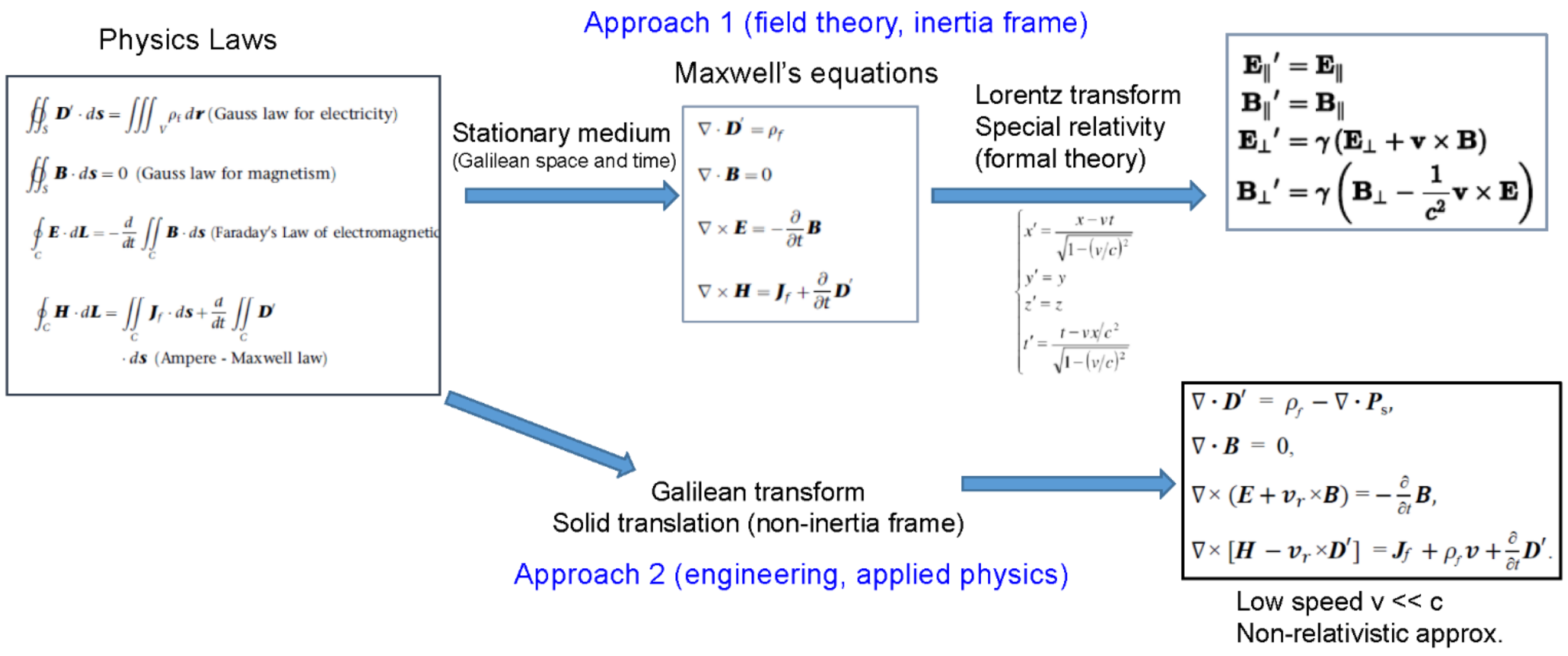

Figure 2. Two fundamental approaches for developing the electrodynamics of a moving media system. The first approach is special relativity through the Lorentz transformation for electromagnetic

phenomena in vacuum space. The second approach is directly derivation from the integral forms of the four physics laws in Galilean space and time, for the case of moving media with specific sizes and shapes.

## 6. Conditions under which the Lorentz transformation approaches the Galilean transformation

In classical electrodynamics and field theory, once it says “motion media”, it usually means a medium that moves at a constant velocity along a straight line. This has been naturally taken as given in almost all text books. In such an inertia system that do not have other form of energy exchange with the outside system, special relativity is most powerful for treating the electromagnetic behavior observed in the Lab frame (at rest) $(\boldsymbol{r}, t)$ ($\boldsymbol{E}$, $\boldsymbol{B}$, $\boldsymbol{D}$, $\boldsymbol{H}$) and the moving frame affixed to the medium $(\boldsymbol{r}', t')$ ($\boldsymbol{E}'$, $\boldsymbol{B}'$, $\boldsymbol{D}'$, $\boldsymbol{H}'$) that is moving at a constant velocity $v_0$ along the x-axis, which are correlated by the Lorentz transformation (Fig. 3a):

$$x' = \gamma(x - v_0 t) \tag{4a}$$

$$y' = y \tag{4b}$$

$$z' = z \tag{4c}$$

$$t' = \gamma(t - x v_0/c^2) \tag{4d}$$

where $\gamma = 1/(1 - v_0^2/c^2)^{1/2}$, and $c$ is the speed of light in vacuum. In the relativistic theory, space and time are unified and correlated. Under the condition of $v_0 \ll c$, the fields in the two reference frames are related by (Fig. 3a):

$$\boldsymbol{E}' \approx \boldsymbol{E} + \boldsymbol{v}_0 \times \boldsymbol{B} \tag{5a}$$

$$\boldsymbol{B}' \approx \boldsymbol{B} - \boldsymbol{v}_0 \times \boldsymbol{E}/c^2 \tag{5b}$$

$$\boldsymbol{D}' \approx \boldsymbol{D} + \boldsymbol{v}_0 \times \boldsymbol{H}/c^2 \tag{5c}$$

$$\boldsymbol{H}' \approx \boldsymbol{H} - \boldsymbol{v}_0 \times \boldsymbol{D} \tag{5d}$$

$$\boldsymbol{J}_f' \approx \boldsymbol{J}_f - \rho_f \boldsymbol{v}_0 \tag{5e}$$

$$\rho_f' \approx \rho_f - \boldsymbol{v}_0 \cdot \boldsymbol{J}/c^2 \tag{5f}$$

In the Galilean transformation:

$$x' = x - v_0 t \tag{6a}$$

$$y' = y \tag{6b}$$

$$z' = z \tag{6c}$$

$$t' = t \tag{6d}$$

where the space and time are absolutely independent. Galilean space and time are most frequently used in classical physics and engineering applications. The fields in the two reference frames are related by:

$$\boldsymbol{E}' = \boldsymbol{E} + \boldsymbol{v}_0 \times \boldsymbol{B} \tag{7a}$$

$$\boldsymbol{B}' = \boldsymbol{B} \tag{7b}$$

$$\boldsymbol{D}' = \boldsymbol{D} \tag{7c}$$

$$\boldsymbol{H}' = \boldsymbol{H} - \boldsymbol{v}_0 \times \boldsymbol{D} \tag{7d}$$

$$\boldsymbol{J}_f' \approx \boldsymbol{J}_f - \rho_f \boldsymbol{v}_0 \tag{7e}$$

$$\rho_f' \approx \rho_f \tag{7f}$$

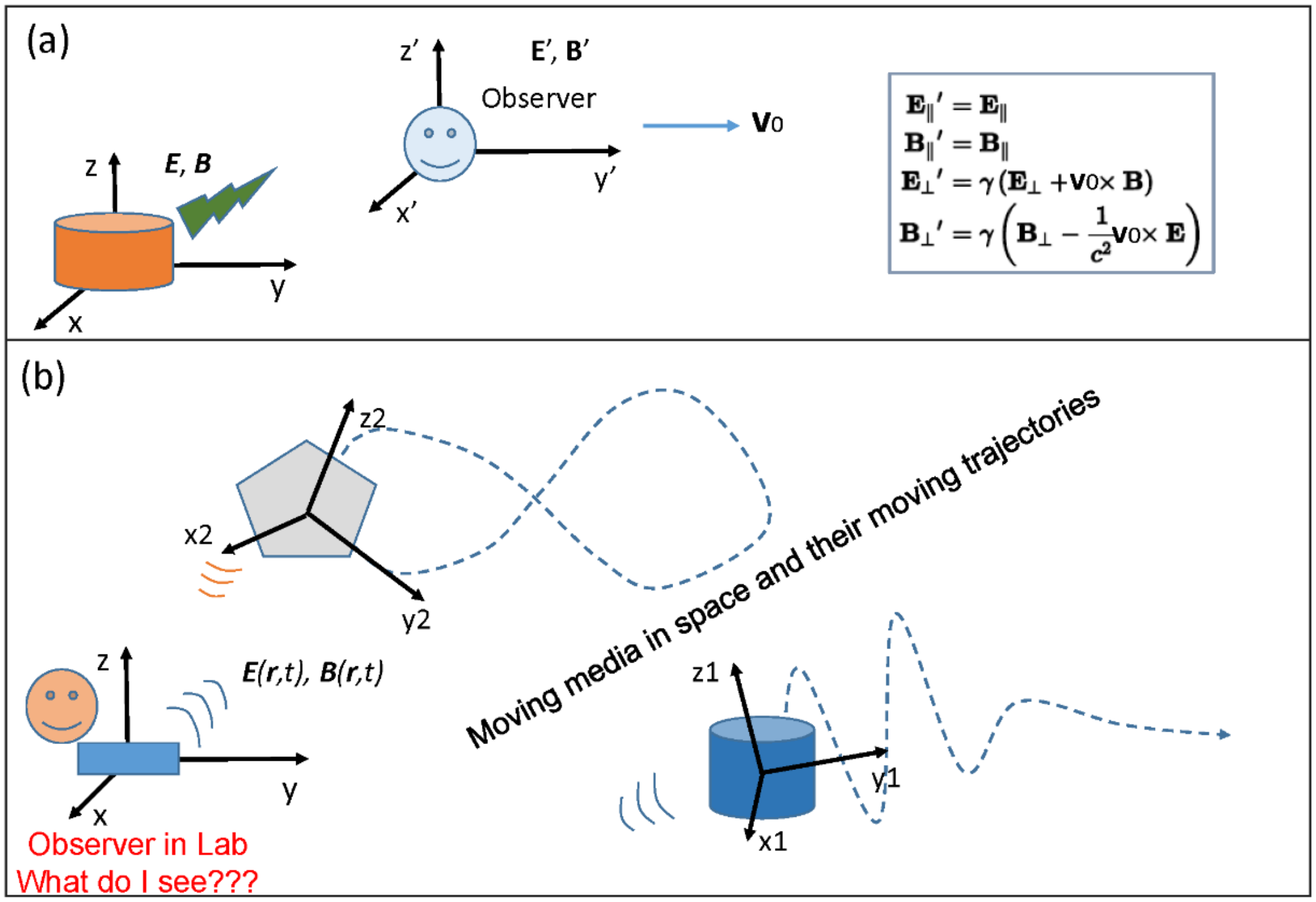

Figure 3. (a) Case for special relativity: Schematic diagram representing the general approach in special relativity, in which the fields observed by an observer who is moving at a constant velocity v about the fields ($\mathbf{E}'$, $\mathbf{B}'$) that was first generated in the Lab frame ($\mathbf{E}$, $\mathbf{B}$). The Lorentz transformed relationships between the fields in the two reference frame in parallel to ***v*** and perpendicular to ***v*** are inset. (b) Mechano-driven moving media system in non-inertia frame: A general case in which the observer is on the ground frame (called Lab frame), with several media moving at complex velocities along various trajectories as represented by the dashed lines. The medium can translate, rotate, expand and even split. The co-moving frames for the media are: (x1, y1, z1), (x2, y2, z2). In such a case, the Lorentz transformation for special relativity cannot be easily applied, and the only realistic approach is to *express all of the fields in the frame where the observation is done (Lab frame) and all of the fields are expressed in the variables in the same frame*.

The conditions under which the Lorentz transformation is equivalent to Galilean transformation are as following [21]:

1. The relative speed between two inertial frames of reference is much smaller than the speed of light in vacuum: $v_0 \ll c$; and
2. Galilean phenomenon takes place in an arena, the spatial extension of which is much smaller than the distance traveled by light during the duration of the phenomenon: $x \ll ct$.

With considering that 9 times of speed of sound is ~3 km/s, which is 1/100,000 of the speed of light, it can be safely stated that for the macroscopic object on earth, the above two conditions are absolutely satisfied, so that Galilean transformation is valid for practical applications. However, in universe, light takes 180 s to travel from earth to Mars, and it takes 100,000 light years to travel across the milky way galaxy. Therefore, the movement of light in universe can be viewed as the moving of a turtle on earth, so slow in comparison to the vast space. Therefore, Lorentz transformation is needed for outer space.

## 7. The Maxwell's equations for a mechano-driven media system that moves with acceleration

Now let's consider the electromagnetic behavior that we have in practical applications on earth. We have two media A and B, each of which has specific shape and size and cannot be regarded as a point, as illustrated in Fig. 1b, both have specific shapes and may move with acceleration one with respect to the other. More importantly, the two may have electromagnetic interaction. Our question is how to describe the electromagnetic behavior of such a system. What we care about is what Bob sees on earth about the two interacting media/phenomena. Such a case is different from the situation for special relativity, which is about how different observers, moving at constant velocity with respect to one another, report their experience of the same physical event. Therefore, our discussions are made based on two assumptions: ignoring relativistic, and $v << c$ (speed of light), which are excellent approximations for almost all of the moving media on earth.

The key concept used in Eq. (1) is the flux: electric flux and magnetic flux! As for the Faraday's law of electromagnetic induction, the changing rate of the *total* magnetic flux through an open surface is the induced electric potential around its closed-edge loop. As for the Ampere-Maxwell's equation, the changing rate of the *total* electric flux through an open surface plus the total current flowing across the surface is the integral of the magnetic field around its closed-edge loop. In the Galilean space and time, we can derive a set of equations by considering the contribution made by the time variation of the media surface to the flux, as shown in Fig. 2 as Approach 2. This approach uses low-speed approximation and ignorance of the relativistic effect. These are excellent approximations for engineering applications on earth [22].

A key mathematical too used for the derivation is the flux theorem: for general functions $g(\boldsymbol{r},t)$ and $\boldsymbol{G}(\boldsymbol{r},t)$, the boundary surface varies with a velocity field $\boldsymbol{v}(\boldsymbol{r},t)$ [12]:

$$\frac{d}{dt}\iint_C \boldsymbol{G}\cdot d\boldsymbol{s} = \iint_C \left[\frac{\partial}{\partial t}\boldsymbol{G} + (\boldsymbol{G}\cdot\nabla)\boldsymbol{v} + (\nabla\cdot\boldsymbol{G})\boldsymbol{v} - (\nabla\cdot\boldsymbol{v})\boldsymbol{G}\right]\cdot d\boldsymbol{s} - \oint_C (\boldsymbol{v}\times\boldsymbol{G})\, d\boldsymbol{L} \tag{8a}$$

$$\frac{d}{dt}\iiint_V g\, d\boldsymbol{r} = \iiint_V \left(\frac{\partial}{\partial t} + \boldsymbol{v}\cdot\nabla\right) g\, d\boldsymbol{r} = \iiint_V \frac{\partial}{\partial t} g\, d\boldsymbol{r} + \oiint_S g\, \boldsymbol{v}\cdot d\boldsymbol{s}. \tag{8b}$$

Using the two mathematical identities as given in Eq. (8), applying them to the integral form of the MEs as given in Eq. (1), and using the Stokes theorem, a set of MEs-f-MDMS is derived [11,12]:

$$\nabla\cdot\boldsymbol{D}' = \rho_f \tag{9a}$$

$$\nabla\cdot\boldsymbol{B} = 0 \tag{9b}$$

$$\nabla\times(\boldsymbol{E} - \boldsymbol{v}\times\boldsymbol{B}) = -\frac{\partial}{\partial t}\boldsymbol{B} \tag{9c}$$

$$\nabla\times(\boldsymbol{H} + \boldsymbol{v}\times\boldsymbol{D}') = \boldsymbol{J}_f + \rho_f\boldsymbol{v} + \frac{\partial}{\partial t}\boldsymbol{D}' \tag{9d}$$

where $\rho_f\,\boldsymbol{v}$ is the current due the medium translation movement, and $\boldsymbol{v}(\boldsymbol{r},t)$ is a function of space and time.

If the movement of the medium is taken as a rigid translation without rotation, $\boldsymbol{v}(t)$ , Eqs. (9a-d) are re-stated as [11]:

$$\nabla\cdot\boldsymbol{D}' = \rho_f \tag{10a}$$

$$\nabla \cdot \boldsymbol{B} = 0 \quad (10b)$$

$$\nabla \times \boldsymbol{E} = -\frac{D}{Dt}\boldsymbol{B} \quad (10c)$$

$$\nabla \times \boldsymbol{H} = \boldsymbol{J}_f + \frac{D}{Dt}\boldsymbol{D}' \quad (10d)$$

where $\frac{D}{Dt} = \frac{\partial}{\partial t} + (\boldsymbol{v} \cdot \nabla)$ (10e)

The law of charge conservation is:

$$\nabla \cdot (\boldsymbol{J}_f + \rho_f \boldsymbol{v}) + \frac{\partial}{\partial t}\rho_f = 0. \quad (10f)$$

where $\rho_f \boldsymbol{v}$ is the current produced by the free charges as the medium being translated at a velocity $\boldsymbol{v}$.

## 8. Expanding the displacement vector for including the mechano-driven polarization

As for the case shown in Fig. 1b, the two media A and B are moving respect to each other, and they may have electrostatic charges on surfaces due to triboelectric or piezoelectric effect. A variation in medium shape and/or moving medium object results in not only a local time-dependent charge density $\rho_s$, but also a local effective electric current density due to the 'passing-by' of the electrostatic charges on the surface of the object once it moves. To account both terms, the displacement vector has to be modified by adding an additional term $\boldsymbol{P}_s$, representing the polarization owing to the pre-existing electrostatic charges on the media, so that the displacement vector $\boldsymbol{D}' = \varepsilon_0 \boldsymbol{E} + \boldsymbol{P}$ has to be replaced by [23, 24]:

$$\boldsymbol{D} = \varepsilon_0 \boldsymbol{E} + \boldsymbol{P} + \boldsymbol{P}_s = \boldsymbol{D}' + \boldsymbol{P}_s \quad (11)$$

Here, the first term $\varepsilon_0 \boldsymbol{E}$ is due to the field created by the free charges, called external electric field; the polarization vector $\boldsymbol{P}$ is the medium polarization caused by the existence of the external electric field $\boldsymbol{E}$; and the added term $\boldsymbol{P}_s$ is mainly due to the existence of the surface electrostatic charges and the medium movement, simply referring to as *mechano-driven polarization*. This term is important for quantifying the output power of TENG.

## 9. Faraday's law of electromagnetic induction for a macroscopic media system that moves at a constant velocity

In order to consider the case for motion with acceleration in a general case, we need to start from the integral forms of the four physics laws, and assume that the media moves at an arbitrary low velocity $\boldsymbol{v}(\boldsymbol{r}, t)$. There are two forms of expressions about the Faraday's law of electromagnetic induction. One is given in Eq. (1c), as used in many text books. The other form is as follows based on the electromotive force [9]:

$$\oint_C \boldsymbol{E}' \cdot d\boldsymbol{L} = -\frac{d}{dt}\iint_C \boldsymbol{B} \cdot d\boldsymbol{s} \tag{12}$$

where $\boldsymbol{E}'$ is the electric field on the moving medium in the frame where $d\boldsymbol{L}$ is at rest. The physics meaning of Eq. (8) is that the total reducing rate of the magnetic flux through an open surface is the work done by the electromotive force ($\xi_{EMF}$) on a unit charge around its closed-edge loop. If the magnetic flux through a surface is $\Phi_B$, the corresponding electromotive force is:

$$\xi_{\mathrm{EMF}} = -\frac{d\Phi_B}{dt} = -\frac{d}{dt}\iint_{\boldsymbol{s}(t)} \boldsymbol{B} \cdot d\boldsymbol{s} \tag{13}$$

This is the flux-rule. The right hand-side of Eq. (12) can be accurately calculated mathematically using the flux theorem Eq. (8a) as follows:

$$\xi_{\mathrm{EMF}} = -\frac{d}{dt}\iint_{\boldsymbol{s}(t)} \boldsymbol{B} \cdot d\boldsymbol{s} = -\iint_{\boldsymbol{s}(t)} \{\frac{\partial}{\partial t}\boldsymbol{B} - \nabla\times[\boldsymbol{v}\times\boldsymbol{B}]\} \cdot d\boldsymbol{s} \tag{14}$$

where $\boldsymbol{v}$ is the velocity at which the boundary surface moves.

Equations (13) and (14) gave two appearing equivalent statements of the Faraday's law of electromagnetic induction, but they are not identical depending on the definition of the looped circuit [25]. Eq. (13) is exact if there is no change in the basic structure of the closed circuit, such as solid connectivity, no relative sliding between the wire and the conductive medium, e.g., the contact between the wires is a "welded" contact. The velocity $\boldsymbol{v}$ in Eq. (14) means the moving velocity of the circuit, which can account for the case in which if there is a relative sliding between the wire and a conductive medium. Therefore, the contact between the wire can be a flexible or changeable contact, such as the case shown in Fig. 4c. Eq. (14) can adequately be applied for the case if the circuit is not a closed loop, such as metal bar moving or rotating in a magnetic field, but Eq. (13) cannot be. More detailed discussions are given in Section 10.

If we consider the case for a moving macroscopic size media, we must start from Eq. (14). Substitution of Eq. (14) into Eq. (12) and use the Stokes theorem, we have

$$\oint_C (\boldsymbol{E}' - \boldsymbol{v}\times\boldsymbol{B}) \cdot d\boldsymbol{L} = -\iint_C \frac{\partial}{\partial t}\boldsymbol{B} \cdot d\boldsymbol{s} \tag{15}$$

This is valid for a general low moving velocity $\boldsymbol{v}$.

The Lorentz force acting on a point charge $q$ in the observer Bob's frame is $\boldsymbol{F} = q\ (\boldsymbol{E}+\boldsymbol{v}_t\times\boldsymbol{B})$, where $\boldsymbol{v}_t$ is the total moving velocity of the point charge that may be different from the moving velocity $\boldsymbol{v}$ of the circuit. Alternatively, the force acting on the same charge in its rest frame is $\boldsymbol{F}' = q\boldsymbol{E}'$. In general, $\boldsymbol{F} \neq \boldsymbol{F}'$ because of the accelerated movement of the circuitry surface boundary, unless the moving frame is *an inertia frame* [26], which means that we must have $\boldsymbol{v} = \boldsymbol{v}_0$. Therefore, only when the medium movement is at a constant speed along a straight line, $\boldsymbol{v} = \boldsymbol{v}_0$, we can have $\boldsymbol{F}' = q\boldsymbol{E}'$ for both inertia reference frames, which gives $\boldsymbol{E}' = \boldsymbol{E}+ \boldsymbol{v}_t\times\boldsymbol{B}$, thus

$$\oint_C [\boldsymbol{E} + (\boldsymbol{v}_t - \boldsymbol{v}_0)\times\boldsymbol{B}] \cdot d\boldsymbol{L} = -\iint_C \frac{\partial}{\partial t}\boldsymbol{B} \cdot d\boldsymbol{s} \qquad (16)$$

Now, the moving velocity of the unit charge can be split into two components: circuit moving velocity $\boldsymbol{v}_0$, and the charge relative velocity $(\boldsymbol{v}_r)$ with respect to the circuit (or medium):

$$\boldsymbol{v}_t = \boldsymbol{v}_0 + \boldsymbol{v}_r \qquad (17)$$

One has to point out that the relative velocity $(\boldsymbol{v}_r)$ of the charge inside a conductor medium may not be small in comparison to the medium moving velocity $\boldsymbol{v}_0$. From Eq. (16),

$$\oint_C [\boldsymbol{E} + \boldsymbol{v}_r \times\boldsymbol{B}] \cdot d\boldsymbol{L} = -\iint_C \frac{\partial}{\partial t}\boldsymbol{B} \cdot d\boldsymbol{s} \qquad (18a)$$

It needs to point out that the moving trajectory of the point charge may not coincidence with the integral path if the medium is moving. Inside the medium, using Stokes theorem and for a general surface:

$$\nabla\times (\boldsymbol{E} + \boldsymbol{v}_r \times\boldsymbol{B}) = -\frac{\partial}{\partial t}\boldsymbol{B} \qquad (18b)$$

Therefore, we must emphasize that the classical mathematical expression of the Faraday's law of electromagnetic induction in differential form as stated in Eq. (3c) can be reduced from Eq. (18a) for a moving media under two conditions:

1. The moving velocity of the medium or circuit boundary is a constant $\boldsymbol{v}_0$ (e.g., inertia frame); and
2. $\boldsymbol{v}_r = 0$, the relative velocity of the unit charge inside the medium is zero; or the medium is a thin wire circuit so that the unit charge is moving parallel to the integral path: $[\boldsymbol{v}_r \times\boldsymbol{B}] \cdot d\boldsymbol{L} = 0$.

Such two conditions were missed in the discussions given by Sheng et al. who naturally assume that the moving velocity of the unit charge is the same as that of the reference frame [27], so that the two terms related to the moving velocity of the reference frame cancels with each other in their derivation. Thus, their conclusion is only valid for the MEs in vacuum space. As for a media that has specific shape and volume, the MEs need to be expanded if there is a medium movement.

There are two possible cases from Eq. (18a):

a) The integral path is a thin wire without intercepting with any large-size conductive medium (Figs. 4a, b), so that the charge moves only along the circuit line, which means $[\boldsymbol{v_r} \times \boldsymbol{B}] \cdot d\boldsymbol{L} = 0$ in Eq. (18a), the Faraday's law is given by:

$$\nabla \times \boldsymbol{E} = -\frac{\partial}{\partial t}\boldsymbol{B} \tag{19}$$

This is the case presented in all of the text books, but it has a condition of thin wire approximation, which needs to be elaborated here. The thin wire circuit is an imaginary circuit whose shape is arbitrary. According to the flux rule, thin wire circuit means that the moving trajectory of the unit charge is exactly the integral path for calculating the magnetic flux, and the circuit is taken as an imaginary circuit when we convert the integral equation into differential form. If the imaginary circuit can be expanded to any space, the thin circuit assumption holds exactly if the circuit does not intercept with any conductive medium boundary, so that it is only exact for vacuum space case, but may not be applicable to the case if there is a moving macroscopic medium.

b) The medium is large so that the relative velocity of the unit charge has a component perpendicular to the integral path in the segment where the integral loop intercepts with a conduction medium, such as the case shown in Fig. 4c, such cases occur in engineering due to the size, shape and volume of the media, which cannot be ignored or taken as a point due to the space in which the electromagnetic behavior is considered.

The example in Fig. 4c can be extrapolated to a general case in which $\oint_C [\boldsymbol{v_r} \times \boldsymbol{B}] \cdot d\boldsymbol{L} \neq 0$ for the space inside the medium, because the charge moving trajectory does not coincidence with that of the integral path, so that the Lorentz force can acts on it locally although there is no change in total magnetic flux from geometrical point of view. This example also gives us an understanding about the meaning of the relative movement velocity $\boldsymbol{v_r}$. Using the Stokes law, for the space inside the medium, Eq. (18a) becomes

$$\nabla \times (\boldsymbol{E} + \boldsymbol{v_r} \times \boldsymbol{B}) = -\frac{\partial}{\partial t}\boldsymbol{B} \tag{20}$$

This case is rarely discussed in classical text books. Eq. (20) can be applied to cases that have a fan-blade shape rotating media and a rotation metal disc in a magnetic field as illustrated in Figs. 4b and c. The finite size of the medium/circuit allows the charges to "wonder" inside the conductive component, so that it is not easy to define where is the integral path of the circuit across the media. This case will be extensively elaborated in Sections 10-11.

This case occurs in TENG because of the use two large platelets as the two electrodes for generating the current, which moves with the dielectric layers beneath in responding to the mechanical driving. TENG relies on a relative movement of the dielectric media and electrodes, so that the "circuit" is an unclosed loop, and its movement is not a "solid" circuit, but a time-dependent, "changeable" structure. Therefore, Eq. (14) has to be used for describing its electromagnetic behavior. The medium movement under external force can be adequately described using Newton mechanics.

## 10. Media motion and Feynman's examples of "anti-flux rule"

Equations (13) and (14) gave two appearing equivalent statements of the Faraday's law of electromagnetic induction, but they are not identical. Eq. (13) is exact if there is no change in the basic structure of the circuit, such as the connectivity, relative sliding between the wire and the conductive medium. The velocity $\boldsymbol{v}$ in Eq. (14) is the moving velocity of the circuit, which can account for if there is a relative sliding between the wire and a conductive medium.

In the textbook by Feynman [20], he outlined a few cases that are "anti-flux rule", as shown in Figs. 4c and d, where the total flux does not change as the fan is rotating in a uniform magnetic field just based on the size of the geometrical area, but there is a potential drop generated along the fan blade due to the Lorentz force (e.g., motion generated potential) [20, 28]. This is because the effective area swiped across by the fan blade as the unit charge travels along the fan blade, which causes a change in magnetic flux. As the fan rotates, the work done by the Lorentz force on the unit charge according to Eq. (14) is: $\xi_{EMF} = \int_0^a r\omega\, B_0\, dr = \frac{1}{2}\,\omega\, a^2\, B_0$ . This result cannot be received if one uses Eq. (13) but Eq. (14).

By expanding the case of Fig. 4b to include a rotating disc in the loop circuit, a potential drop and electric current are generated across the load R as the disc rotates, although the total flux through the closed loop does not change with time. This seems against the total flux rule (Eq. (1c)), but it can be understood as follows. In this case, the charge can move inside the conductive disc at a velocity of $v_r =$ $r\omega$ perpendicular to the radial direction, where $\omega$ is the angular velocity, thus the total electromotive force is along the entire radius $a$ from $P_1$ to O, with a potential drop calculated using Lorentz force:

$$\xi_{EMF} = \int_a^0 [\boldsymbol{v}_r \times \boldsymbol{B}] \cdot d\boldsymbol{r} = \int_0^a r\omega\, B_0\, dr = \frac{1}{2}\, \omega\, a^2\, B_0\, , \tag{21}$$

and the current flowing in the circuit is $\frac{1}{2R}\ \omega\, a^2\, B_0$. Alternatively, using the flux rule (Eq. 13)), the area of the disc segment between the two dashed lines is: $A = \frac{1}{2}\ \omega t\, a^2$, so according to the flux rule (Eq. (13)), $\xi_{EMF} = AB_0/t = \frac{1}{2}\ \omega\, a^2\, B_0$. The results are consistent with that of Eq. (20) using Lorentz force.

The case shown in Fig. 4c was attributed by Feynman as an "anti-flux rule" example. This paradox is likely due to that the path of the unit charge moving in the disc (blue dashed line) deviates from the original rectangular "circuit" (as indicated by black dashed line in Fig. 4c-I), along which the integral for calculating the magnetic flux is done. As the charge enters the disc at point $P_1$ at its edge at $t = 0$, it moves along the radial to point $P_2$ as the disc rotates at $t = t$ (Fig. 4c-II); its moving path is indicated by a blue dashed line. Therefore, the area as defined by the two dashed lines in Fig. 4c-II is the effective area of change of magnetic flux. *This change in flux is due to the deviation of the unit charge transport path from that of the geometrical path as the disc rotates. Therefore, the understanding of flux rule needs to consider the charge transport process if there is a relative movement in the conductive medium and a change in circuit structure, such as sliding.*

The "anti-flux rule" case occurs in Fig. 4c is also due to that the accelerated moving media is a conductor. If the moving medium is an insulator, the situation could be different, because there is no conductive circuit that defines the integral loop. Therefore, one may need to consider the conductivity of the media case by case.

The example illustrated in Fig. 4c is a great example regarding to electrodynamics of a moving media system. The thin wire section is at stationary, but the metal disc is rotating, with a needle sliding on it edge. The moving velocities of the space points inside the disc are space (r) and time dependent, and there is acceleration during disc rotation. The electrodynamics in this system has to be treated using Eq. (20) rather than Eq. (19)! This is a good example of why we have to expand the Maxwell's equations for a moving media system.

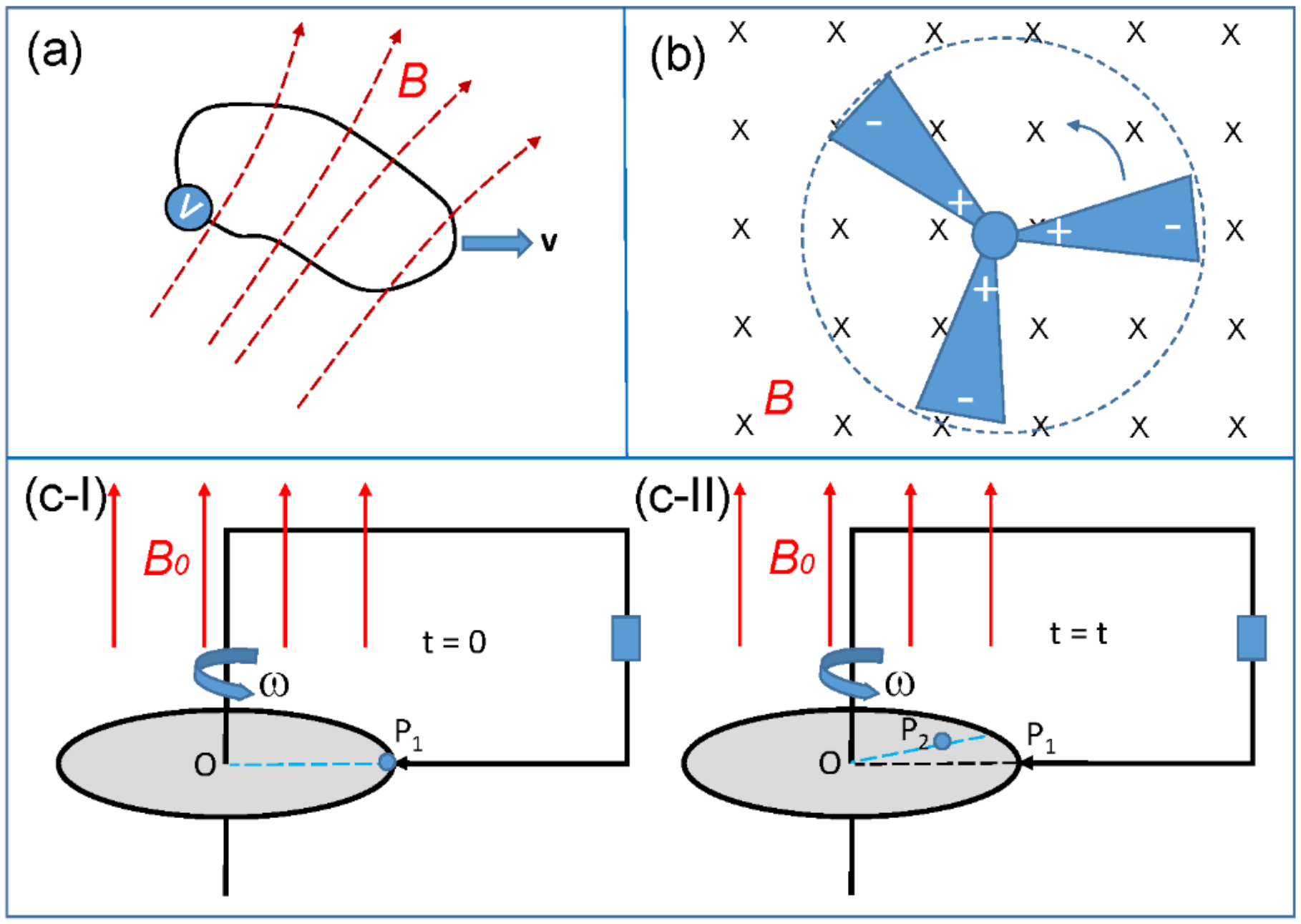

Figure 4. Several cases regarding to the flux rule for electromagnetic induction. (a): A typical example of thin wire circuit that is moving with acceleration with respect to the Lab frame in a time- and space-dependent magnetic field. (b): A conductive fan that is rotating inside a uniform magnetic field. (c) A rectangular thin wire circuit that is at stationary in a uniform magnetic field, but with one end sliding on the edge of a rotating conductive disc at *t = 0* and *t = t*, and the other end is connected to the axis of the disc. The unit charge within the macroscopic size object can move, which leads to an additional term in the Faraday's law for electromagnetic induction.

## 11. Faraday's law of electromagnetic induction for a macroscopic media system that moves with acceleration

If the medium is moving with an acceleration, so that the inertia force may have to be considered once we consider the total force acting on the unit charge by standing on the rest reference frame of the circuit. In such a case, the force acting on a unit charge $q$ has to include the inertia force, we may have:

$$q\boldsymbol{E}' - \frac{\partial}{\partial t}(m\boldsymbol{v}) = q\boldsymbol{E} + q\boldsymbol{v}_t \times \boldsymbol{B} \tag{22}$$

where $m$ is the total mass of the unit charge, the amount of which remain to be determined. If the unit charge is taken as a virtual charge that may have no mass, but just for theoretical calculation, the mass related term vanishes; however, we are not sure if this is the case. The Faraday's law may be written as:

$$\oint_C \left[\boldsymbol{E} + \boldsymbol{v}_r \times \boldsymbol{B} + \frac{1}{q}\frac{\partial}{\partial t}(\mathrm{m}\boldsymbol{v})\right] \cdot d\boldsymbol{L} = -\iint_C \frac{\partial}{\partial t}\boldsymbol{B} \cdot d\boldsymbol{s} \tag{23}$$

where the moving velocity $\boldsymbol{v}$ of the media is arbitrary as long as $v \ll c$. We now discuss the following cases.

a) If the integral path is a looped thin wire circuit that does not intercept with a large conductive plate, so that the relative moving velocity of the charge is parallel to the integral path, the second term at the left-hand side of Eq. (23) vanishes. The second condition is that the loop moving velocity $\boldsymbol{v}$ is a rigid translation, $\boldsymbol{v}(\mathrm{t})$, $[\nabla\times(\mathrm{m}\boldsymbol{v})] = 0$, the third term at the left-hand side of Eq. (23) vanishes as well, we have the classical expression of Faraday's law:

$$\nabla\times \boldsymbol{E} = -\frac{\partial}{\partial t}\boldsymbol{B} \tag{24}$$

b) If the integral path is a circuit that intercepts with a large medium, so that the relative moving velocity of the charge may not be parallel to the integral path within the conductive medium, the second term at the left-hand side of Eq. (23) remains. The second condition is that the loop moving velocity $\boldsymbol{v}$ is a rigid translation, $\boldsymbol{v}(\mathrm{t})$, $[\nabla\times(\mathrm{m}\boldsymbol{v})] = 0$, the third term at the left-hand side of Eq. (23) vanishes as well, we have the Faraday's law in in its expanded format:

$$\nabla\times (\boldsymbol{E} + \boldsymbol{v}_r\times\boldsymbol{B}) = -\frac{\partial}{\partial t}\boldsymbol{B} \tag{25}$$

c) For a general case in which the circuit loop has a time dependent and position dependent moving velocity, $\boldsymbol{v}(\mathbf{r},t)$, such as for liquid and elastic media, the term $\frac{\partial}{\partial t}[\nabla\times(\mathrm{m}\boldsymbol{v})]$ is related to the rotation movement of the loop circuit. The accelerated rotation motion of a shape-deformable circuit (e.g., a liquid or elastic loop, or expandable loop) can produce an electric field. Therefore, the term $\frac{\partial}{\partial t}[\nabla\times(\mathrm{m}\boldsymbol{v})]$ characterizes a "source" of electromagnetic radiation owing to the accelerated rotation of the "expandable/flexible" loop circuit.

Alternatively, if one assumes that the unit charge is a virtual charge that has no mass, the last term in Eq. (23) also vanishes.

## 12. Expansion of the Ampere-Maxwell's law for a macroscopic media system that moves with acceleration

The expansion of the Ampere-Maxwell's law to the cases of large medium that moves with an acceleration is not straight forward. The reason is that the displacement current was first introduced in the differential form of the MEs in order to satisfy the law of conservation of charges, so the term $\frac{\partial}{\partial t}\boldsymbol{D}$ was added in the Ampere's law by Maxwell for such a purpose. Therefore, the physical meaning of $\oint_C \boldsymbol{H} \cdot d\boldsymbol{L}$ is not as straight forward as that for Faraday's law using the concept of electromotive force. With considering the symmetry between electricity and magnetism as well as the equivalence of the two fields, the Ampere-Maxwell's law for the space inside a medium may be equivalent written as:

$$\nabla \times (\boldsymbol{H} - \boldsymbol{v_r} \times \boldsymbol{D}) = \boldsymbol{J_f} + \rho_f \boldsymbol{v} + \frac{\partial}{\partial t}\boldsymbol{D} \qquad (26)$$

Summary all of the discussions presented above, the MEs-f-MDMS for a media system that moves as a rigid translation $\boldsymbol{v}(\mathrm{t})$, the electrodynamics inside the media are described by:

$$\nabla \cdot \boldsymbol{D'} = \rho_f - \nabla \cdot \boldsymbol{P}_s, \qquad (27a)$$

$$\nabla \cdot \boldsymbol{B} = 0, \qquad (27b)$$

$$\nabla \times (\boldsymbol{E} + \boldsymbol{v_r} \times \boldsymbol{B}) = -\frac{\partial}{\partial t}\boldsymbol{B}, \qquad (27c)$$

$$\nabla \times [\boldsymbol{H} - \boldsymbol{v_r} \times (\boldsymbol{D'} + \boldsymbol{P}_s)] = \boldsymbol{J_f} + \rho_f \boldsymbol{v} + \frac{\partial}{\partial t}[\boldsymbol{D'} + \boldsymbol{P}_s]. \qquad (27d)$$

Equations (23a-d) are the MEs-f-MDMS for a media system that moves with an arbitrary but low velocity even with acceleration, with the inclusion of motion induced mechano-polarization, which describes the coupling among three fields: mehcano – electricity – magnetism. Take the case shown in Fig. 4c as an example, $\boldsymbol{v_r} = r\omega\hat{\varphi}$ within the disk. If there is no relative movement of the unit charge with respect to the medium, $\boldsymbol{v_r} = 0$, Eqs. (27a-d) resumes the expression for classical MEs.

For engineering applications, our observation is done in the Laboratory frame on earth (Fig. 5). If there are several moving media in space, the electrodynamics inside each media is governed by the MEs-f-MDMS. In this case, one does not have to worry about the medium speed exceeding that of the speed of light in vacuum $c_0$, because the speed of the light inside media $c_m$ is always slower than $c_0$, and $v_r \ll c_m$. The electrodynamics outside the medium in vacuum space is governed by the classical MEs, which means that the speed of light remains constant regardless the media is moving or not. The solutions of two sets of equations meet at the media boundaries as governed by boundary conditions.

The derived MEs-f-MDMS does not conflict with that of the MEs for field theory. The former is for the space inside of a moving medium, while the letter is for the vacuum space, in which all of the media are treated as point charges. The two sets of equations govern two different zones. As the field theory is mostly about universe, so that there is no worry from the theoretical physics point of view.

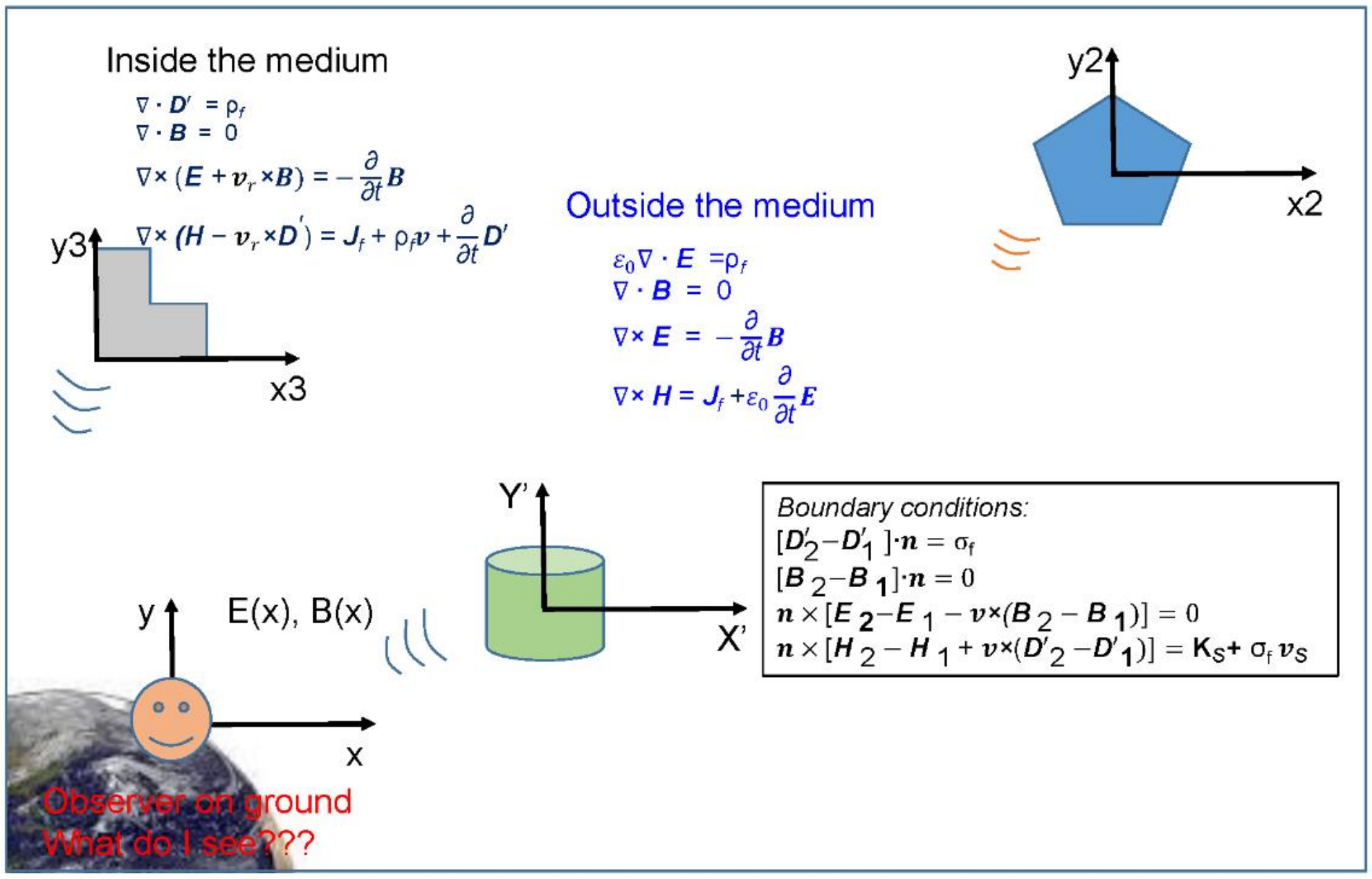

Figure 5. Conjunction of the Maxwell equations for a mechano-driven system inside the media and the classical Maxwell equations in the vacuum space, as the media are moving in space, and the observation is done on earth (in Lab frame).

## 13. Conservation of energy in the mechano-electric-magnetic coupling system

Starting from Eqs. (23a-d), we derived the energy conservation process in this mechano-electric-magnetic coupled process [12]:

$$-\frac{\partial}{\partial t}u - \nabla\cdot\boldsymbol{S} = \boldsymbol{E}\cdot\boldsymbol{J}_f + \rho_f\boldsymbol{v}\cdot\boldsymbol{E} + \{\boldsymbol{H}\cdot[\nabla\times(\boldsymbol{v}_r\times\boldsymbol{B})] + \boldsymbol{E}\cdot[\nabla\times(\boldsymbol{v}_r\times(\boldsymbol{D}'+\boldsymbol{P}_s))]\}. \tag{28}$$

where $\boldsymbol{S}$ is the Poynting vector

$$\boldsymbol{S} = \boldsymbol{E}\times\boldsymbol{H}, \tag{29}$$

and $u$ is the energy volume density of electromagnetic field, which is generally given by

$$\frac{\partial}{\partial t}u = \boldsymbol{E}\cdot\frac{\partial\boldsymbol{D}}{\partial t} + \boldsymbol{H}\cdot\frac{\partial\boldsymbol{B}}{\partial t}. \tag{30}$$

This equation means that the decrease of the internal electromagnetic field energy within a volume plus the rate of electromagnetic wave energy radiated out of the volume surface is the rate of energy done by the field on the external free current and the free charges, plus the media spatial motion induced change in electromagnetic energy density. The contribution made by media movement can be considered as "sources" for producing electromagnetic waves! This is what we have been observed experimentally [15,16,17,18,19].

If we assume that the speed term depends only on time, $\boldsymbol{v}(t)$, it can be simplified as

$$-\frac{\mathrm{D}}{\mathrm{D}t}u - \nabla\cdot\boldsymbol{S} = \boldsymbol{E}\cdot\boldsymbol{J}_f, \tag{31a}$$

with $\frac{\mathrm{D}}{\mathrm{D}t}u = \boldsymbol{E}\cdot\frac{\mathrm{D}\boldsymbol{D}}{\mathrm{D}t} + \boldsymbol{H}\cdot\frac{\mathrm{D}\boldsymbol{B}}{\mathrm{D}t}$, (31b)

and $\frac{D}{Dt} = \frac{\partial}{\partial t} - (\boldsymbol{v}_r \cdot \nabla)$ (31c)

Note, $\boldsymbol{v}_r$ is the velocity of the unit charge with respect to the media instead of the moving velocity of the medium.

## 14. Maxwell equations for a mechano-driven system in tensor format

If we can approximately use the conventional constitutive relationship for isotropic media: $\boldsymbol{B} = \mu\boldsymbol{H}$, and assume that the movement of the media is a rigid translation, the Maxwell's equations for a mechano-driven system is expressed in the classical format for field theory:

$$\varepsilon\nabla\cdot\boldsymbol{E} = \rho' \tag{32a}$$

$$\nabla\cdot\boldsymbol{B} = 0 \tag{32b}$$

$$\nabla\times\boldsymbol{E} = -\frac{D}{Dt}\boldsymbol{B} \tag{32c}$$

$$\nabla\times\boldsymbol{B} = \mu\boldsymbol{J}' + \mu\varepsilon\frac{D}{Dt}\boldsymbol{E} \tag{32d}$$

The law of conservation of charges is

$$\nabla\cdot\boldsymbol{J}' + \frac{D}{Dt}\rho_f = 0. \tag{32e}$$

where:

$$\rho' = \rho_f - \nabla\cdot\boldsymbol{P}_s \tag{33a}$$

$$\boldsymbol{J}' = \boldsymbol{J}_f + \rho_f\boldsymbol{v}_t + \frac{D}{Dt}\boldsymbol{P}_s \tag{33b}$$

We can use the definition of the vector potential and scalar potential:

$$\boldsymbol{B} = \nabla\times\boldsymbol{A} \tag{34a}$$

$$\boldsymbol{E} = -\nabla\varphi - \frac{D}{Dt}\boldsymbol{A} \tag{34b}$$

Substitute Eqs. (34a-b) into Eq. (33a-d), we have,

$$\nabla^2 \boldsymbol{A} - \mu\varepsilon \frac{D^2}{Dt^2} \boldsymbol{A} = -\mu \boldsymbol{J}' \quad (35a)$$

$$\nabla^2 \varphi - \mu\varepsilon \frac{D^2}{Dt^2} \varphi = -\frac{\rho'}{\varepsilon} \quad (35b)$$

Under condition:

$$\nabla \cdot \boldsymbol{A} + \mu\varepsilon \frac{D}{Dt} \varphi = 0 \quad (35c)$$

where

$$\frac{D^2}{Dt^2} = [\frac{\partial}{\partial t} - (\boldsymbol{v}_r \cdot \nabla)][\frac{\partial}{\partial t} - (\boldsymbol{v}_r \cdot \nabla)] = \frac{\partial^2}{\partial t^2} - (\boldsymbol{v}_r \cdot \nabla)\frac{\partial}{\partial t} - \frac{\partial}{\partial t}(\boldsymbol{v}_r \cdot \nabla) + (\boldsymbol{v}_r \cdot \nabla)(\boldsymbol{v}_r \cdot \nabla) \quad (35d)$$

The term $\frac{\partial}{\partial t}\boldsymbol{v}_r$ is the acceleration, which represents the applied external force.

We now transform Eqs. (32a-e) into the format of tensors, and the Maxwell's equations for a mechano-driven system is expressed in the classical format for field theory. We now use the classical expressions of following quantities for electrodynamics, the anti-symmetric strength tensor of electromagnetic field,

$$F^{\alpha\beta} = \xi^{\alpha} A^{\beta} - \xi^{\beta} A^{\alpha} \quad (36a)$$

$$F_{\alpha\beta} = \xi_{\alpha} A_{\beta} - \xi_{\beta} A_{\alpha} \quad (36b)$$

where α, β = (1,2,3,4), and the newly defined operators are

$$\xi^{\alpha} = (\frac{1}{c}\frac{D}{Dt}, -\boldsymbol{\nabla}) \quad (37a)$$

$$\xi_{\alpha} = (\frac{1}{c}\frac{D}{Dt}, \boldsymbol{\nabla}) \quad (37b)$$

$$A^{\alpha} = (c\varphi, \boldsymbol{A}) \quad (37c)$$

$$A_{\alpha} = (c\varphi, -\boldsymbol{A}) \quad (37d)$$

One can prove

$$F^{\alpha\beta} = \begin{pmatrix} 0 & -E_x/c & -E_y/c & -E_z/c \\ E_x/c & 0 & -B_z & B_y \\ E_y/c & B_z & 0 & -B_x \\ E_z/c & -B_y & B_x & 0 \end{pmatrix} \quad (38a)$$

$$F_{\alpha\beta} = \begin{pmatrix} 0 & E_x/c & E_y/c & E_z/c \\ -E_x/c & 0 & -B_z & B_y \\ -E_y/c & B_z & 0 & -B_x \\ -E_z/c & -B_y & B_x & 0 \end{pmatrix} \quad (38b)$$

where $c = c_m = 1/(\mu\varepsilon)^{1/2}$. For an easy exercise, one can easily prove

$$E_x = -\frac{\partial}{\partial x}\varphi - \frac{D}{Dt}A_x = -(\xi^0 A^1 - \xi^1 A^0) \quad (39a)$$

$$B_x = \frac{\partial}{\partial y}A_z - \frac{\partial}{\partial z}A_y = -(\xi^2 A^3 - \xi^3 A^2) \quad (39b)$$

Eqs. (32a-e) can be restated as [**Error! Bookmark not defined.**]:

$$\xi_\alpha F^{\alpha\beta} = \mu J^\beta \quad (40)$$

where $J^\beta = (c\rho', \boldsymbol{J}')$. This is the Maxwell's equations for a mechano-driven system. This is the Maxwell's equations for a mechano-driven system. Note Eq. (40) is the same as that for the classical Maxwell's equations except the operator $\partial_\alpha$ is replace by $\xi_\alpha$.

If $\boldsymbol{v}_r$ = 0, Eq. (40) resumes the form of classical Maxwell's equations.

## 15. The Lagrangian of the Maxwell's equations for a mechano-driven system

We now derive the Lagrangian $L$ for the Maxwell's equations for a mechano-driven system. Λ is assumed to be a function of the density of the Lagrangian of the system $\Lambda(A_\alpha, \xi_\alpha A_\beta)$. We vary the action

$$\delta\int_{-\infty}^{\infty} L\, dt = \delta\iint_{-\infty}^{\infty} \Lambda(A_\alpha, \xi_\alpha A_\beta)\, d\boldsymbol{r}dt = 0 \quad (41)$$

which gives

$$\iint_{-\infty}^{\infty} [\frac{\partial\Lambda}{\partial A_\alpha}\delta A_\alpha + \frac{\partial\Lambda}{\partial(\xi_\alpha A_\beta)}\delta(\xi_\alpha A_\beta)]\, d\boldsymbol{r}dt = 0 \quad (42)$$

Now we look at the second term and integrate by part over (ct, x, y, z) [e.g. ($x_0$, $x_1$, $x_2$, $x_3$)], respectively, with considering the vanishing of the function at infinity. If the medium motion is a rigid translation $\delta\boldsymbol{v}(t)$ that is only time dependent, we have

$$\iint_{-\infty}^{\infty}[\frac{\partial \Lambda}{\partial A_\alpha}\delta A_\alpha]\, d\boldsymbol{r}dt - \iint_{-\infty}^{\infty}\{\, \xi_\alpha \left[\frac{\partial \Lambda}{\partial(\xi_\alpha A_\beta)}\right] \delta A_\beta\}\, d\boldsymbol{r}dt \tag{43}$$

We have the Lagrangian relation:

$$\frac{\partial \Lambda}{\partial A_\beta} - \xi_\alpha \frac{\partial \Lambda}{\partial(\xi_\alpha A_\beta)} = 0 \tag{44}$$

The density of the Lagrangian for the electromagnetic field is given by [29]

$$\Lambda = F^{\alpha\beta} F_{\alpha\beta} + \mu J^\alpha A_\alpha \tag{45}$$

Substituting Eq. (45) into Eq. (44), we have the Maxwell's equations for a mechano-driven system

$$\xi_\alpha F^{\alpha\beta} = \mu J^\beta \tag{46}$$

## 16. Electrodynamics of a moving media system in non-inertia reference frame – an approximated approach

As presented above, the classical mathematical expression of the *Faraday electromagnetic induction law in differential form holds for moving media only if its moving velocity is a constant and along a straight line*! e.g., in inertia frame! By the same token, we also believe that the mathematical expressions for the differential form of Ampere-Maxwell law may holds only for inertial frame. This concept can be generalized as: the differential form of the Maxwell's equations may hold only in inertia frame, and it is subject to be expanded for non-inertia frame/motion [11, 12].

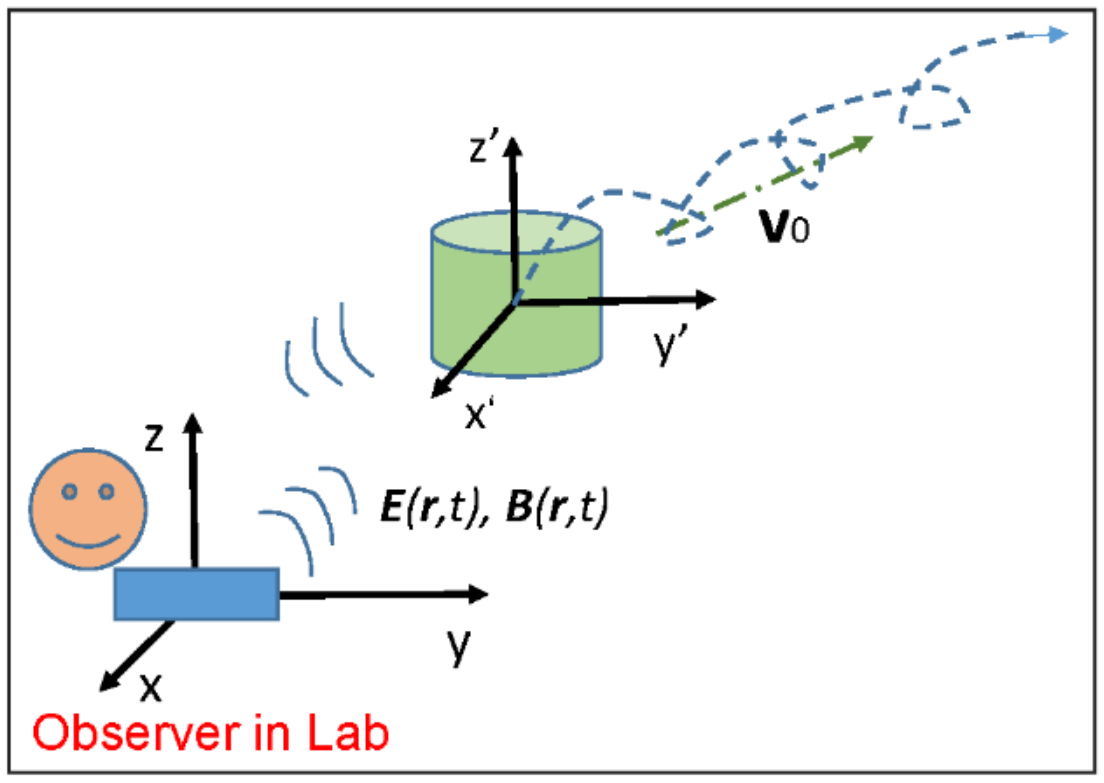

Figure 6. Schematic diagram showing an approximated method for decompose a non-inertia movement as a movement in an inertia frame plus a correction in non-inertia frame.

If the moving velocity of the medium is a constant for inertial frame , the relationship can be easily derived from special relativity under slow speed limit as $\boldsymbol{E}' = \boldsymbol{E} + \boldsymbol{v}_0 \times \boldsymbol{B}$. But for non-inertial frame, the media moves at an arbitrary low velocity with acceleration, we can adopt an approximate approach, in which the motion velocity is split into a constant component $\boldsymbol{v}_0$ and a time-dependent component $\delta\boldsymbol{v}$,

$$\boldsymbol{v}(\boldsymbol{r},t) = \boldsymbol{v}_0 + \delta\boldsymbol{v}(\boldsymbol{r},t) \tag{47}$$

where the term $\delta\boldsymbol{v}(\boldsymbol{r},t)$ contains both rotation and small component of rigid translation. If the local field can be approximately treated as $\boldsymbol{E}' \approx \boldsymbol{E} + \boldsymbol{v}_0 \times \boldsymbol{B}$ by only considering the constant velocity component, which is schematically illustrated in Fig. 6. The idea is that we try to use the results developed for an inertial reference frame for approximately treating the case in a non-inertia reference frame,

$$\oint_C \boldsymbol{E}' \cdot d\boldsymbol{L} \approx \oint_C (\boldsymbol{E} + \boldsymbol{v}_0 \times \boldsymbol{B}) \cdot d\boldsymbol{L} = \iint_{\boldsymbol{s}(t)} \nabla \times [\boldsymbol{E} + \boldsymbol{v}_0 \times \boldsymbol{B}] \cdot d\boldsymbol{s} \tag{48}$$

This equation means that the changing rate of the flux of the magnetic field through an open surface plus the work done by the Lorentz force due to media movement on a unit charge around its edge loop is the circulation of the induced electromotive force around its closed edge loop. Therefore:

$$\nabla \times (\boldsymbol{E} - \delta\boldsymbol{v} \times \mathrm{B}) \approx -\frac{\partial}{\partial t}\boldsymbol{B} \tag{49}$$

By the same token, in analogy to the Faraday's electromagnetic induction law, we can start from the integral form of the Ampere-Maxwell law in the same fashion:

$$\oint_C \boldsymbol{H}' \cdot d\boldsymbol{L} = \iint_C \boldsymbol{J}_f \cdot d\boldsymbol{s} + \frac{d}{dt}\iint_C \boldsymbol{D} \cdot d\boldsymbol{s} \tag{50}$$

where the $\boldsymbol{H}'$ is the magnetic field on the moving medium in the frame where $d\boldsymbol{L}$ is at rest. Using the flux theorem, we have

$$\oint_C \boldsymbol{H}' \cdot d\boldsymbol{L} = \iint_C (\boldsymbol{J}_f + \rho_f \boldsymbol{v}) \cdot d\boldsymbol{s} + \iint_C \frac{\partial}{\partial t}\boldsymbol{D} \cdot d\boldsymbol{s} - \oint_C (\boldsymbol{v} \times \boldsymbol{D}) \cdot d\boldsymbol{L} \tag{51}$$

This equation means that the sum of, the total current flowing through an open surface and the current produced by the free charges due to media movement (right-hand side first term), the changing rate of the electric flux through the open surface (right-hand side second term), and the electric circulation produced by media movement (right-hand side third term), is the circulation of the magnetic field around its closed edge loop. Using the Stokes theorem, Eq. (51) becomes

$$\nabla\times(\boldsymbol{H}' + \boldsymbol{v}\times\boldsymbol{D}) = \boldsymbol{J}_f + \rho_f\boldsymbol{v} + \frac{\partial}{\partial t}\boldsymbol{D} \quad (52)$$

Note, $\boldsymbol{H}'$ is the magnetic field in the moving frame affixed to the media that moves at an arbitrary low velocity $\boldsymbol{v}$.

Under low speed limit and in inertia frame, the local magnetic field in the moving frame is $\boldsymbol{H}' \approx \boldsymbol{H} - \boldsymbol{v}_0\times\boldsymbol{D}$, we have

$$\nabla\times(\boldsymbol{H} + \delta\boldsymbol{v}\times\boldsymbol{D}) \approx \boldsymbol{J}_f + \rho_f\boldsymbol{v} + \frac{\partial}{\partial t}\boldsymbol{D} \quad (53)$$

The terms of $\delta\boldsymbol{v}\times\boldsymbol{B}$ and $\delta\boldsymbol{v}\times\boldsymbol{D}'$ are the sources of produced electromagnetic waves by media movement. In the case of the moving velocity is constant, $\boldsymbol{v}(\boldsymbol{r},t) = \boldsymbol{v}_0,\ \delta\boldsymbol{v} = 0$, Eqs. (49, 53) resume that of the Classical Maxwell equations.

The choice of a constant moving velocity as the basic reference frame allow us to introduce the approximated constitutive relations for non-inertia frame. If we approximately use the constitutive relations derived for constant velocity of motion:

$$\boldsymbol{D} \approx \varepsilon\boldsymbol{E} - \varepsilon\boldsymbol{v}_0\times\boldsymbol{B} \quad (54a)$$

$$\boldsymbol{H} \approx \boldsymbol{B}/\mu + \varepsilon\boldsymbol{v}_0\times\boldsymbol{E} \quad (54b)$$

the Maxwell equations for a general case can be stated as follows:

$$\varepsilon\nabla\cdot\boldsymbol{E} = \rho_f + \varepsilon\nabla\cdot(\boldsymbol{v}_0\times\boldsymbol{B}) \quad (55a)$$

$$\nabla\cdot\boldsymbol{B} = 0 \quad (55b)$$

$$\nabla\times(\boldsymbol{E} - \delta\boldsymbol{v}\times\boldsymbol{B}) = -\frac{\partial}{\partial t}\boldsymbol{B} \quad (55c)$$

$$\nabla\times(\boldsymbol{B}/\mu + \varepsilon\boldsymbol{v}_0\times\boldsymbol{E} + \varepsilon\delta\boldsymbol{v}\times\boldsymbol{E}) = \boldsymbol{J}_f + \rho_f\boldsymbol{v} + \frac{\partial}{\partial t}(\varepsilon\boldsymbol{E} - \varepsilon\boldsymbol{v}_0\times\boldsymbol{B})$$

$$\approx \boldsymbol{J}_f + \rho_f\boldsymbol{v} + \varepsilon\frac{\partial}{\partial t}\boldsymbol{E} + \varepsilon\boldsymbol{v}_0\times(\nabla\times\boldsymbol{E})$$

Since

$$\nabla\times(\boldsymbol{v}_0\times\boldsymbol{E}) = \boldsymbol{v}_0(\nabla\cdot\boldsymbol{E}) - (\boldsymbol{v}_0\cdot\nabla)\boldsymbol{E}$$

$$\nabla(\boldsymbol{v}_0\cdot \boldsymbol{E}) = (\boldsymbol{v}_0\cdot \nabla)\boldsymbol{E} + \boldsymbol{v}_0 \times (\nabla\times \boldsymbol{E})$$

$$\boldsymbol{v}_0 \times (\nabla\times \boldsymbol{E}) = \nabla(\boldsymbol{v}_0\cdot \boldsymbol{E}) - (\boldsymbol{v}_0\cdot \nabla)\boldsymbol{E} = \nabla(\boldsymbol{v}_0\cdot \boldsymbol{E}) + \nabla\times (\boldsymbol{v}_0 \times \boldsymbol{E}) - \rho_f \boldsymbol{v}_0/\varepsilon$$

We have

$$\nabla\times (\boldsymbol{B}/\mu + \varepsilon\delta\boldsymbol{v}\times \boldsymbol{E}) = \boldsymbol{J}_f + \rho_f\, \delta\boldsymbol{v} + \varepsilon\, \nabla(\boldsymbol{v}_0\cdot \boldsymbol{E}) + \varepsilon\frac{\partial}{\partial t}\boldsymbol{E} \tag{55d}$$

It must be pointed out that the quantities in Eqs. (55a-d) derived directly from the integral Maxwell's equations are given in Lab frame (***r***, *t*). Equs. (55a-d) now have (***E***, ***B***) as the variables and they can be solved approximately using the classical method.

## 17. Discussions

There are two fundamental understandings about space and time: relativistic space and time, and absolute space and time. Therefore, there are two approaches for deal with the electrodynamics of moving media (Fig. 2). If the moving velocity is uniform in inertia frame, special relativity can be easily applied to this case without assuming low-moving speed limit. In this system, the total energy of electricity and magnetism is conservative. In special relativity, the general approach is starting from the integrated form of the four physics laws, a set of differential equations is derived (Maxwell's equations) for stationary media whose shape and boundary are independent of time. Then, the electromagnetic behavior of a moving media to be observed in Lab is described using the Lorentz transformation. Such coordination transformation is taken as the formal approach for moving media system. However, for a media system that moves in non-inertial frame, which means that the speed is a function of time at least, the theory for general relativity may be required for such a case, which is probably too complicated to be used for engineering purposes. For the speed we care about and the engineering purpose on earth, we believe that there is no need to use special relativity for applied electromagnetism.

In approach 1 (Fig. 2), if we start using the Galilean transformation instead of the Lorentz transformation, we could get the results of the Galilean electromagnetism [30], in which the field in space at a time t is taken as a quasi-static case, e.g. the "frozen" field assumption. Therefore, the media distribution and related fields are treated "frame by frame" (as in films for a movie) under quasi-static approximation. The theory of moving media can be treated frame by frame with the use of constitutive relationships under slow-media moving cases. More approximations can be made for magnetic-dominated or electric-dominated systems. Again, such theory can be easily applied if the moving speed is a constant along a straight line, but we have to check if it can be applied for complex media moving trajectory cases, because we are not sure if the constitutive relationship derived using special relativity would hold in non-inertial frame.

The second approach in Fig. 2 is to directly starting from the four physics laws by deriving all of the fields in the Lab frame and in Lab coordination system using the Galilean space and time without making a coordination transformation [6,11,12]. The most important advantage of this approach is that it can be applied to any media that move along complex trajectories in non-inertial frame as long as the moving speed is low and the relativistic effect is ignored. Such equations should not be Lorentz covariant simply due to the energy input from mechanical triggering and accelerated media motion. This approach is more effective for applied physics, which has been widely used in engineering electrodynamics [31, 32].

Lastly, we have introduced the extended form of the Faraday's law with considering the relative moving velocity of the charge with the presence of a large conductive medium and even with the presence of the inertia force produced by the movement of the reference frame. In all of the text books, the third equation in the Maxwell's equations is for a case that the conductive medium is a linear circuit, which is exact for vacuum case. However, if the conductive medium is a plane or sheets in which the unit charge can deviate from the integral path, an additional term has to be included in the Faraday's law for electromagnetic induction. More rigorously, if the inertia force is included and if it is significant, another term has to be included as well. Our discussion does indicate that the Faraday's law of electromagnetic induction needs to be expanded for media that move with an acceleration, although the terms introduced may be small.

We need to point out that there are two formats of the integral equations for the Faraday's electromagnetic induction law: Eq. (1c) and Eq. (12). Equation (12) is for field theory, and it is built based on the electromotive force model. Eq. (1c) is more often used for engineering electrodynamics. We believe that the results received from both statements would be equivalent if one keeps consistent within its own theoretical frame and as long as we are confined to the engineering applications on earth.

## 18. Conclusions and perspectives

One of the key questions that has been debated for is: if the MEs should be expanded if the media system is moving. Our first conclusion is that, if the described space is in vacuum, there is no need to expand the MEs regardless if there are media or not in the nearby. Inside the media that have finite sizes and volumes, the newly derived MEs-f-MDMS may be required for describing the local electrodynamics of the medium that moves with and even without acceleration. This is because the finite size and volume of the medium could lead to the movement of the unit charge within the volume that may not exactly follow the integral path when calculating the magnetic flux, especially with the existence of conductive plates/discs. Outside of the moving media and in the vacuum space, the local electrodynamics is governed by the classical MEs. The total solutions of the two sets of equations meet at the boundaries. Therefore, one can simply conclude that, for the electrodynamics of macroscopic objects on earth for engineering applications, one may need to use the MEs-f-MDMS if there is medium movement. Lastly, the theory is presented in Lagragian scheme, from which the Maxwell's equations for a mechano-driven media is derived.

The history of developing the electromagnetism is a great example to learn (Fig. 7). Since the first discovery of the electromagnetic induction phenomenon in 1831 by Faraday, the mathematical description of the physics phenome using the flux concept was accomplished by Lenz in 1834 based on Faraday's many experimental observations. Based on the existing laws of electricity and magnetism, Maxwell established the first set of MEs in 1861 by mathematically introduced the displacement current. In comparison to conduction current, displacement is a "current" that transmit through space at the speed of light. The current version of MEs was only finalized by about 1900 after the major contributions from several scientists. The prediction of electromagnetic wave was only on theory until Hertz verified it experimentally in 1888. Although this was one of the most exciting discoveries as we see it from now, but as the electromagnetic wave was first discovered, Hertz said "It's of no use whatsoever. This is just an experiment that proves Maestro Maxwell was right. We just have these mysterious electromagnetic waves that we cannot see with our naked eye. But they are there". What Hertz did not realize was that this was the beginning of wireless communication. As today, the entire world is covered by wireless communication.

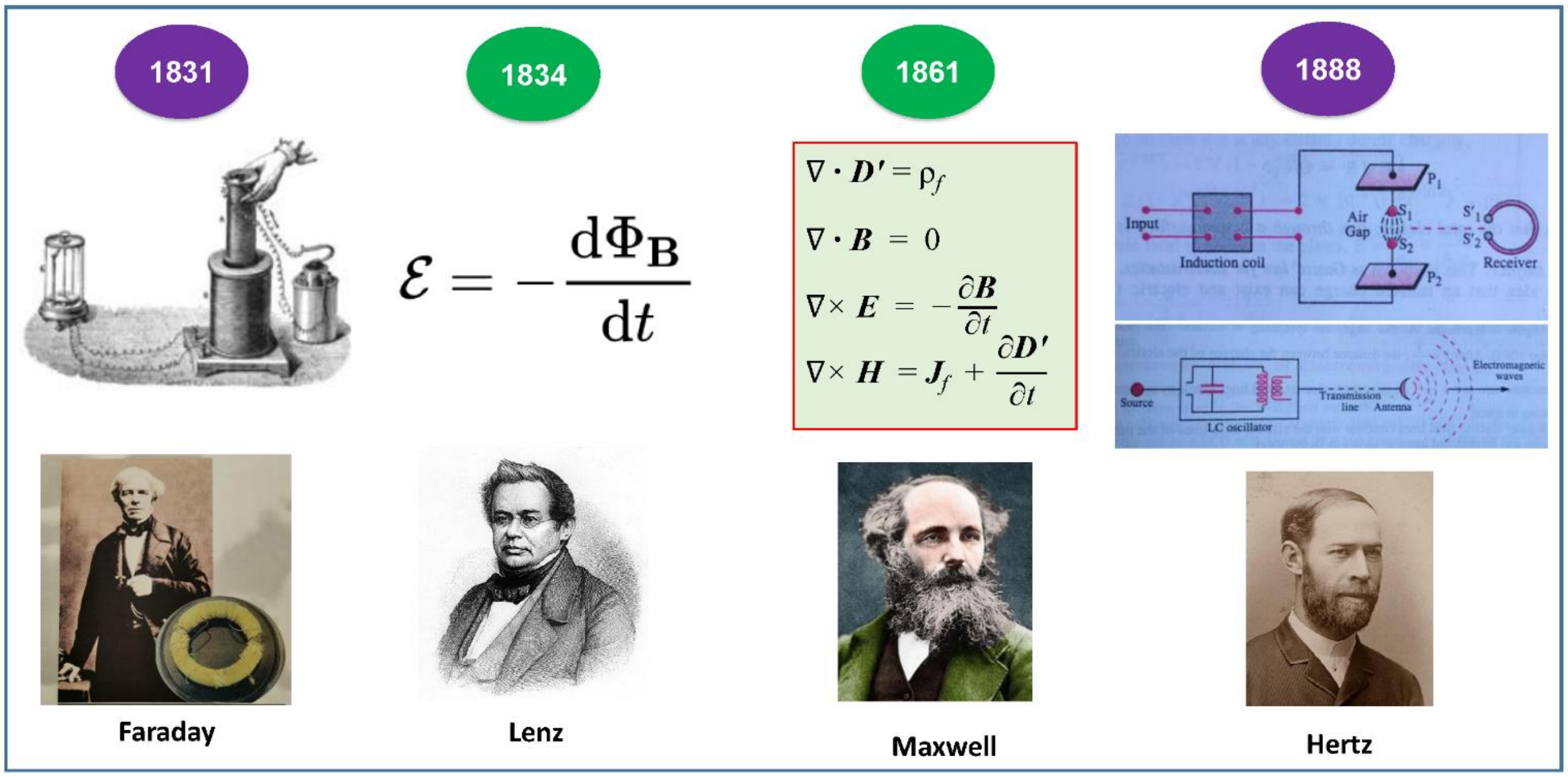

Figure 7. A brief history regarding to the development of electrodynamics.

As an inspiration, now let's look forward regarding to the introduction of the MEs-f-MDMS (Fig. 8). We experimentally discovered the first piezoelectric nanogenerator in 2006 and the triboelectric nanogenerators in 2012 based on simple experimental meausrements, we first introduced the mechano-driven polarization $\boldsymbol{P}_s$ term in the MEs in 2017 for quantifying the output of TENG. Then, we systematically expanded the MEs in 2021. Although we have had a few sets of experiments that proved the necessities for introducing the MEs-f-MDMS, we should be open minded regarding to its future potential applications. The current research is built based on new experimental observations that were unavailable at Maxwell's time, and most of text books on electrodynamics were written based on the experimental observations at least 70-100 years ago. With considering the new experimental observations to meet our current technological demands, it is necessary to expand the classical electrodynamics to the situations that experiments can reach.

To be inspired, we quote the prediction about wireless technology by Tesla in 1926 way before the birth of transistors and integrate circuit: "When wireless is perfectly applied the whole earth will be coverted into a huge bran, which in fact it is, all things being particles of a real and rhythmic whole. We shall be able to communicate with one another instantly, irrespective of distance. Not only this, but through television and telephone we shall see and hear one another as perfectly as though we were face to face, despite intervening distances of thousands of miles; and the instruments through which we shall be able to do this (fit in a) vest pocket". Of course, to reach a dream "the best way to predict the future is to create it", said by Lincoln.

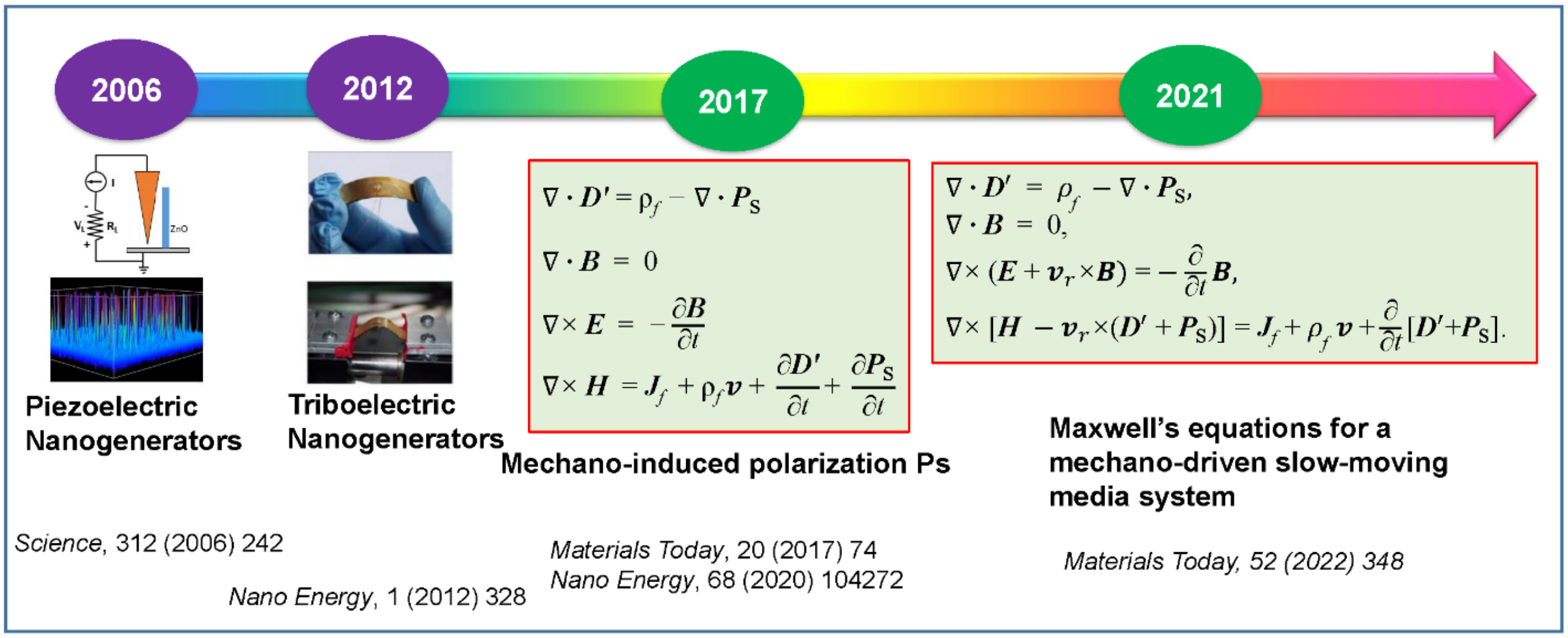

Figure 8. Our journey regarding the expansion of the Maxwell's equations for a mechano-driven slow moving medium system. The future remains to be discovered.